\documentstyle[12pt]{article}
\begin{document}
\thispagestyle{empty}
\bibliographystyle{unsrt}
\newcommand\bea{\begin{equation}}
\newcommand\eea{\end{equation}}
\vskip 2cm
\begin {center} {\large \bf ANNIHILATORS OF IRREDUCIBLE MODULES
AND KINEMATICAL CONSTRAINTS OF PAIR OPERATORS}
\end {center}
\vskip 1cm
\centerline{{\large  Mircea Iosifescu }
\footnote {E-mail address: iosifesc@theor1.ifa.ro}}
\vskip 0.3 cm
\centerline {Department of Theoretical Physics}
\centerline { Institute for Atomic Physics, P.O.Box MG-6,
Bucharest, Romania}
\vskip 3cm \nopagebreak \begin {abstract}
The kinematical constraints of pair operators in nuclear collective motion,
 pointed out by Yamamura and identified by Nishiyama as relations
 between so(2n) generators, are recognized as equations
satisfied by second-degree annihilators (deduced in previous work)  
of irreducible so(2n)-modules. The recursion relations for Nishiyama's 
 tensors and their dependence on
 the parity of the tensor degree is explained. An explanation is
 also given for the recursion relations for sp(2n) tensors 
 pointed out by Hwa and Nuyts. The statements
 for the algebras so(2n) and sp(2n) are proved simultaneously.  

\end {abstract} 
\newpage 
\newpage \setcounter{page} 1
\section {Introduction} \par

In his algebraic approach to the theory of nuclear collective
motion, Yamamura \cite{yam} pointed out a number of polynomial kinematical
constraints that have to be satisfied by the pair operators 
(which are defined as commutators of fermions).

Observing that the pair operators generate the Lie algebra
so(2n), Nishiyama derived \cite{nish} these constraints by using
the commutation relations of the generators of this algebra.
He constructed, recurrently, sets of homogeneous polynomials of
increasing degrees in the so(2n) generators, each set possessing
a specific so(2n) tensor property which depends only on the
parity of the degree of the polynomials which compose the set. The
second-degree polynomials and the recursion relations are then
used in \cite{nish} to derive Yamamura's kinematical
constraints, whose Lie algebraic nature is established in this way.

Similar recursion relations had been pointed out by Hwa and
Nuyts in their construction of sp(2n) generators as
anticommutators of boson operators \cite{hwa}. Their study had,
however, not the aim to investigate any identity satisfied by
the sp(2n) generators.

Let $\Lambda_{i},~(i=1,...,n)$ be the fundamental
weights of a Lie algebra of rank $n$.

The present paper points out that Nishiyama's polynomials of second
degree are precisely the irreducible tensors of highest weight $2\Lambda_{1}$
in the universal enveloping algebra U(so(2n)) of the so(2n) algebra,
whose explicit expressions have been derived in previous work
\cite{is85}-\cite{is88-2} by projection from a generic 
second-degree element. These tensors were proved \cite{is88-2}
to be annihilators of so(2n)-modules which transform under (spinorial)
representations of highest weights $k\Lambda_{n-1}$ and $k\Lambda_{n}$
of so(2n).

As it is known, (e.g. \cite{miller}), the spinorial
representations of so(2n) can be constructed in terms of the
generators of the Clifford algebra (in other words, in terms
of creation and annihilation operators which anticommute);
Nishiyama's result, which has been derived in a constructive
way, appears thus as a consequence of the property of the $(2\Lambda_{1})$ 
tensors in $U(so(2n))$ to be annihilators of spinorial modules.

The present paper proves also in a simple way 
that the tensors constructed by recursion in \cite{nish}  
belong to irreducible representations (IRs) that depend only on the
parity of the tensor degree. The similar property observed in \cite{hwa}
for sp(2n) is also proved.

Simultaneous proofs have been given for this property for the so(2n)
and sp(2n) algebras. Indeed, as the sp(2n) (so(2n)) generators can be
constructed as anticommutators (commutators) of boson (fermion) operators,
the two cases can be distinguished only by the value of a parameter
$\epsilon$ ($\epsilon = +1$ for so(2n) and $\epsilon = -1$ for sp(2n)):
we obtain in this way analogous expressions for the irreducible
tensors and a simultaneous proof for the recurrence relation.

The identification of Nishiyama's kinematical constraints as
second-degree tensors in the universal enveloping algebra stresses
the physical relevance of the equations which result from the
vanishing of these tensors, whose importance for physics has been already
pointed out in
\cite{is88-1},\cite{is88-2},\cite{is80}-\cite{is90}.
 
\vskip 1cm
\section {Irreducible tensors in U(L) and S(L)}

\vskip .3cm
Physicists knew since a long time that specific realizations of a
Lie algebra L satisfy specific  sets of relations
 (for a partial list of references cf. \cite{is88-2})
 each relation consisting in the vanishing of a well-defined homogeneous
polynomial whose indeterminates are the generators of the realization.

As observed in \cite{is85},\cite{is88-2} a s~et of such
polynomials has specific tensorial properties under the adjoint
action.

 A realization $\rho$ of a Lie algebra L is 
a homomorphism $\rho: L \rightarrow A$ of L into an
associative algebra A endowed with a Lie bracket $[~,~]_{A}$, 
compatible with associativity. The homomorphism $\rho$ can be extended to a 
homomorphism of associative algebras $\rho_{U}: U(L) \rightarrow
A$ or $\rho_{S}: S(L) \rightarrow A $ (where $U(L)$ and $S(L)$
are the universal enveloping algebra of L and the symmetric algebra of L,
 respectively);
extensions of the adjoint action of $L$ to
$S(L)$ and to $U(L)$ can be also defined, as well as irreducible tensors
under this extension.

The irreducible tensors $T_{\Lambda}$ (of highest weight
$\Lambda$) of the extension of $ ad L $ to $U(L)$ transform under
subrepresentations $(\Lambda)$ of the symmetric Kronecker powers
$(ad^{\otimes k})_{sym}$ of $ad L$. (This is a consequence of
the property of the symmetrization operator to intertwine
tensors in $S(L)$ and $U(L)$). As pointed out elsewhere \cite{is88-2},
the tensors $T_{\Lambda}$ vanish on specific irreducible
representations $(\lambda)$ of $ L $, i.e. $T_{\Lambda}$ are
annihilators of the $L$-module $(\lambda)$.

 For classical semisimple Lie algebras all second-degree tensors
$T_{\Lambda}$ in $S(L)$ that transform under a subrepresentation 
(of highest weight $\Lambda$) of the symmetric part $(ad \otimes
ad)_{sym}$ of the Kronecker square of the adjoint representation
have been determined \cite{is85}: the corresponding tensors
$T_{\Lambda}$ in the second-degree component $U^{2}(L)$ of 
$U(L)$ result from the tensors in $S^{2}(L)$ by
symmetrization with respect to the order of the factors in each product.

  For the algebras $A_{n}, B_{n}, C_{n}$ and $D_{n}$
 all finite-dimensional L-modules $(\lambda)$ which are annihilated by 
irreducible second-degree tensors $T_{\Lambda}$ in $U(L)$  have
 been determined in \cite{is88-2} for all second-degree symmetric 
tensors $T_{\Lambda}$.
\vskip .1cm

The irreducible tensors $T_{\Lambda}$ into which decomposes the second-degree
component $S^{(2)}$ of $S(L)$ result by projection 
$P_{i}S_{\alpha\beta}S_{\gamma\delta}$
from a generic element $S_{\alpha\beta}S_{\gamma\delta}\in S^{2}(L)$,
where $P_{i}$ is
the projection operator associated with the eigenvalue $c_{i}$
of the Casimir operator $C$ of the IR $\Lambda$:
\bea P_{i} = \frac{\prod_{j\neq i}(C-c_{j}I)}{\prod_{j\neq i}
(c_{j}-c_{i})} \eea
 This calculus will be performed 
simultaneously for the algebras $sp(2n,C)$ and $o(2n,C)$, for
which identical $\epsilon$-dependent expressions can be derived for
the generators, the Casimir elements, the projection operators and,
finally, for the irreducible tensors.
\vskip 2cm
{\bf The generators of o(2n,C) and sp(2n,C)}
\vskip .1cm

Let us remind some well-known results.

The o(2n,C) (sp(2n,C)) Lie algebra is defined as the set of
linear operators $\Gamma$, acting in a 2n-dimensional complex vector
space, which leaves invariant a non-degenerate symmetric
(antisymmetric) bilinear form 
$~(u,v) = \epsilon (v,u)~$
with $~\epsilon~= +1~~(\epsilon = -1)~$ for the orthogonal (symplectic)
algebra; the invariance of $~(u,v)~$ under an element $\Gamma$ of the Lie 
algebra means that $(\Gamma u,v)+(u,\Gamma v)=0$
 
In order to treat the o(2n,C) and sp(2n,C) symmetry in a similar
way one chooses for the symmetric bilinear form the expression
\bea
(u,v) = u^{t}Kv
\eea \\
where $X^{t} $ means the transpose of $X$ and $K$ is the $2n\times 2n$ matrix
\bea
 K = \left (\matrix {0&I_{n} \cr \epsilon I_{n}&0 \cr} \right)
\eea
( $I_{n}$ is the n-dimensional unit matrix).
Introducing the coordinates  $~u_{j},~v_{j}~~(j=1,...,2n)~$, the
bilinear form $(u,v)$ becomes 
\bea
(u,v)= \sum_{i=1}^{n}(u_{i}v_{i+n}+ \epsilon u_{i+n}v_{i})
\eea 
 The invariance property of the bilinear form $(u,v)$ leads to
the following unique condition to be satisfied by the matrices of
 the orthogonal and symplectic algebras
\bea \Gamma^{t}K+K\Gamma =0 \eea
with K given by (3); decomposing the matrix $\Gamma$ into $n \times n$ blocks
 \bea \Gamma = \left(\matrix{A&B\cr -C&D\cr}\right) \eea  
 we get from Eq. (5) the following conditions for the $n \times n$ matrices
A, B, C and D :
\bea
B=-\epsilon B^{t}~,~ C=-\epsilon C^{t}~,~ D=-A^{t}
\eea

The generators $\Gamma$ of the o(2n,C) and sp(2n,C)
 algebras are thus expressed as linear combinations of the 
following basic elements:
\bea
e_{ij}-e_{j+n,i+n}~~ for~ the~ generators~ belonging~ to~~A \oplus (-A^{t})
\eea
\bea
e_{i,j+n}-\epsilon e_{j,i+n}~~~ for~ the~ generators~ belonging~ to~~~ B 
\eea
\bea
e_{i+n,j}-\epsilon e_{j+n,i}~~~ for~ the~ generators~ belonging~ to~~~ -C
\eea
where  $~i,j=1,...,n~$  and  $~e_{\alpha\beta}~$  are elements of the
 Weyl basis  $~(e_{\alpha\beta})_{\gamma\delta} =
 \delta _{\alpha\gamma} \delta _{\beta\delta}~$ with 
$1\leq \gamma,\delta\leq 2n$.

The generators of the two algebras can be written in the
following form valid for both algebras
\bea
S_{\alpha \beta} = \sum_{\lambda =1}^{2n} (g_{\alpha \lambda}e_{\lambda
\beta}-g_{\lambda \beta}e_{\lambda \alpha})~~~~(\alpha,\beta=1,...,2n)
\eea
where  
\bea g_{\alpha \beta } = \delta _{\alpha, \beta +n} + \epsilon \delta
_{\alpha+n,\beta }~~~,~~~ g_{\alpha \beta} = \epsilon g_{\beta \alpha}~~~,~~~
 S_{\alpha \beta} = -\epsilon S_{\beta \alpha}        \eea

Let us observe that $g\equiv(g_{\alpha,\beta}) =K^{-1}$
and that $~g^{\alpha\beta}\equiv (g^{-1})_{\alpha\beta}=
\epsilon g_{\alpha\beta}$.

With the notations (11)-(12), the commutation relations have then
the same expression for both algebras, namely
\bea
[S_{\alpha \beta},S_{\gamma \delta}] = g_{\gamma \beta}S_{\alpha
\delta}+g_{\delta \alpha}S_{\beta \gamma}-g_{\alpha
\gamma}S_{\beta \delta}-g_{\beta \delta}S_{\alpha \gamma}
~~~  (\alpha,\beta,\gamma,\delta = 1,...,2n) \eea

The $n \times n$ blocks  $A, B, C, D$  are recovered from
Eqs.(11) and (12)
by reintroducing the labels  $~i,j = 1,...,n$; 
we get:
\bea S_{ij} = \epsilon e_{i+n,j}-e_{j+n,i} = -\epsilon S_{ji} \eea
\bea S_{i+n,j+n}= e_{i,j+n}-\epsilon e_{j, i+n} = -\epsilon S_{j+n,i+n}\eea
\bea S_{i+n,j}= e_{i,j}-e_{j+n,i+n}=-\epsilon S_{j,i+n} \eea
and comparison with the expressions (8-10) leads to the identifications: 
\bea S_{i+n,j} = (A\oplus (-A^{t}))_{i,j} \eea
\bea S_{i+n,j+n} = B_{i,j}  \eea
\bea S_{i,j} =-\epsilon C_{i,j}  \eea
Denoting, for simplicity, by $A$ the sum of the matrices $A
\oplus -A^{t}$,  i.e denoting  $A_{i,j} = S_{i+n,j}$,  the unique
commutation relation (13) is equivalent with the
following set of distinct commutation relations $(i,j,k,l \leq n)$:
\bea [A_{ij},A_{kl}] = \delta _{jk} A_{il} - \delta _{il}
A_{kj} \eea
\bea [B_{ij},B_{kl}] = [C_{ij},C_{kl}] = 0 \eea
\bea [A_{ij}, B_{kl}]=\delta_{jk} B_{il}-\epsilon \delta _{jl}B_{ik}
\eea
\bea [A_{ij},C_{kl}]=\epsilon \delta _{il}C_{jk}-\delta
_{ik}C_{jl} \eea
\bea [B_{ij},C_{kl}]=-\delta _{jk}A_{il}-
\delta _{il}A_{jk}+\epsilon \delta _{ik}A_{jl}+\epsilon\delta_{jl}A_{ik} \eea
\newpage
{\bf Killing-Cartan bilinear form and dual elements.}.
\vskip .3cm
The expression (11) of the generators leads to the
following expression of the Killing-Cartan form valid both for so(2n,C) and
sp(2n,C) 
\bea (S_{\alpha \beta},S_{\gamma \delta}) = tr(adS_{\alpha
\beta}adS_{\gamma \delta}) = 
8n (g_{\alpha \delta}g_{\gamma \beta} -
 \epsilon g_{\alpha \gamma}g_{\beta \delta}) \eea
i.e. taking into account Eqs.(13)
\bea (S_{\alpha \beta},S_{\gamma \delta})=
8n[(\delta_{\alpha,\delta+n}+\epsilon \delta_{\alpha+n,\delta})(\delta
_{\gamma,\beta+n}+\epsilon \delta _{\gamma+n,\beta})-
(\delta_{\alpha,\gamma+n}+\epsilon
\delta _{\alpha+n,\gamma})(\delta _{\beta,\delta+n}+
\epsilon \delta _{\beta+n,\delta})] \eea 
whence the following pairs of elements which are dual with
respect to the Killing-Cartan form can be determined:

\bea(S_{\alpha \leq n,\beta\leq n}, \epsilon S_{\beta+n,\alpha+n}),~~~
(S_{\alpha>n,\beta>n},\epsilon S_{\beta-n,\alpha-n}), \eea
\bea(S_{\alpha\leq n,\beta>n},S_{\beta-n,\alpha+n}),~~~
(S_{\alpha>n,\beta\leq n},S_{\beta+n,\alpha-n})   \eea
or, using the notation introduced by the decomposition in blocks:

\bea dual~of~B_{ij} = \epsilon C_{ji},~~~
     dual~of~C_{ij} = \epsilon B_{ji},~~~
     dual~of~A_{ij} = A_{ji} \eea
\vskip.3cm    

{\bf Casimir element and projection operators}
\vskip .3cm
Let $\rho:L\rightarrow gl(V)$ be a representation of the Lie
algebra L and $\beta$ a bilinear symmetric form associated with
 $\rho$; let $~x_{i}$ and $x^{i}~~(i=1,...,dimL)~$ be dual elements of L
 with respect to $\rho$. The {\em Casimir
element} of the representation $\rho$ 
\bea C_{\rho}(\beta) =
\sum_{i=1}^{dimL}\rho(x_{i})\rho(x^{i}) \eea
commutes with $\rho$ and is basis independent. 
The action of the Casimir element $C=C_{ad}$ of the adjoint representation 
on the element $S_{\alpha\beta}$ writes

$~~ CS_{\alpha\beta} =
\sum_{\kappa,\lambda,\mu,\nu=1}^{2n}g_{\kappa\nu}g_{\mu\lambda}
[S_{\kappa\lambda},[S_{\mu\nu},.]] =  8(n-\epsilon)S_{\alpha\beta}~~ $.
   Using the identity
$$ [A,[B,CD]]=[A,C][B,D]+[B,C][A,D]+C[A,[B,D]]+[A,[B,C]]D $$
we write the action $C^{k}S_{\alpha\beta}S_{\gamma\delta}$
of the powers $C^{k}$ of the Casimir element on the product
$S_{\alpha,\beta}S_{\gamma,\delta}$ of two generic elements 
and calculate the expression of the projecion operator (1).

The action of the projection operators $P_{\Lambda_{2}}$ for
$L=sp(2n),~(\epsilon=-1)$ and $(P_{2\Lambda_{1}})$ for $L=o(2n),
~(\epsilon=+1)$,
on the product $S_{\alpha\beta}S_{\gamma\delta}$ gives tensors
of highest weights $\Lambda_{2}$ and $2\Lambda_{1}$, respectively:
\vskip.1cm

$$ P_{\Lambda_{2}(2\Lambda_{1})}S_{\alpha\beta}S_{\gamma\delta} =
 \frac{C(C-8(n-2\epsilon))(C-4(2n-\epsilon))}{(-4n)(8n-16\epsilon-4n)
(8n-4\epsilon-4n)}S_{\alpha\beta}S_{\gamma\delta} =$$
$$ \frac{1}{2n-2\epsilon}[g_{\alpha\gamma}(T_{\beta\delta}-\frac{1}{2n}g_{\beta
\delta}I_{2})+g_{\beta\delta}(T_{\alpha\gamma}-
\frac{1}{2n}g_{\alpha\gamma}I_{2})-$$
\bea~~~~~~~~~~~~~~~~~~~~~~~~~~~~\epsilon g_{\alpha\delta}
(T_{\beta\gamma}-\frac{1}{2n}g_{\beta\gamma}I_{2})-\epsilon 
g_{\beta\gamma}
(T_{\alpha\delta}-\frac{1}{2n}g_{\alpha\delta}I_{2})]~~~~~~~~~~~~~~\eea
where

\bea T_{\alpha\beta}=S_{\alpha\lambda}g^{\lambda\mu}S_{\mu\beta}
\eea
and
\bea I_{2}=
g_{\alpha\beta}g_{\gamma\delta}S_{\alpha\delta}S_{\gamma\beta} 
\eea
is the second-degre invariant.
$T_{\alpha\beta}$ are components of the irreducible tensor in $S(L)$  
with highest weight $\Lambda_{2}$ for $sp(2n)$
and $2\Lambda_{1}$ for $o(2n)$.

The vanishing of the elementary components in Eq.(31) lead to the equations:
\bea E_{\alpha,\beta}\equiv T_{\alpha\beta} - 
\frac{1}{2n}g_{\alpha\beta}I_{2}=0  \eea
Reminding the decomposition (7) in $n\times n$ blocks and the
defining equations (11),(17-19), we obtain that Eq.(34) is 
equivalent with the following four sets of equations (in which 
$~i,j=1,2,...,n)$ :
\bea E_{i,j}\equiv T_{ij}=(AB-BA^{t})_{ij}=0   \eea
\bea E_{i+n,j+n}\equiv T_{i+n,j+n}=-(CA-A^{t}C)_{ij}=0  \eea
\bea E_{i,j+n}\equiv (A^{2}-BC)_{ij}-\frac{\delta_{ij}}{2n}I_{2}=0  \eea
\bea E_{i+n,j}\equiv \epsilon((A^{t})^{2}-CB)_{ij}-
\frac{\delta_{ij}}{2n}I_{2})=0 \eea
where the expression of the Casimir invariant $I_{2}$ is:
\bea I_{2}= tr(A^{2}-BC+(A^{t})^{2}-CB) \eea
The corresponding tensors in $U(L)$ are obtained by symmetrization 
of Eqs.(35-38) with respect to order:
\bea (AB-BA^{t})_{ij}+\epsilon (AB-BA^{t})^{t})_{ij}=0  \eea
\bea (CA-A^{t}C)_{ij}+\epsilon (CA-A^{t}C)^{t})_{ij}=0  \eea
\bea (A^{2}-BC)_{ij}+((A^{t})^{2}-CB)^{t}_{ij}=
2\frac{\delta_{ij}}{2n}I_{2} \eea
 It has been proved \cite{is88-2} that these equations
are verified by the generators of the IRs $(m\Lambda_{n})$
for $sp(2n)$ and $(m\Lambda_{n-1}),~(m\Lambda_{n})$ for $so(2n)$,
i.e. that the components of these tensors are annihilators of the 
irreducible modules $(m\Lambda_{n})$ for $sp(2n)$ and $(m\Lambda_{n-1})$,
$(m\Lambda_{n})$ for $o(2n)$; for the $o(2n)$ algebra the
corresponding IRs are spinorial.
\newpage

\section {The Hwa-Nuyts-Nishiyama (HNN) realisations}
\vskip .2cm

We shall treat the HNN realisations and recursion relations 
on an equal footing
and consider therefore a Fock space $F_{\epsilon}$ for a system
with n degrees of freedom on which representations of the
canonical commutation $~(\epsilon = -1)~$ or anticommutation 
$~(\epsilon = +1)~$ relations are defined:
\bea [b_{i},b_{j}^{+}]_{\epsilon}\equiv b_{i}b_{j}^{+}+\epsilon
b_{j}^{+}b_{i} = \delta_{ij}I  \eea
\bea [b_{i},b_{j}]_{\epsilon} = [b_{i}^{+},b_{j}^{+}]=\delta_{ij}I~~~
(i,j=1,...,n)  \eea

Realisations of $sp(2n,R)$ have been defined on $F_{-1}$ by Hwa and
Nuyts \cite{hwa}; realisations of $o(2n,R)$ have been defined on $F_{+1}$  
by Nishiyama \cite{nish}. Both realisations can be written as
\bea E^{i}_{j}\equiv  \frac{1}{2}[b_{i}^{+},b_{j}]_{-\epsilon} =
\frac{1}{2}(b_{i}^{+}b_{j}-\epsilon b_{j}b_{i}^{+}) = 
b_{i}^{+}b_{j}-\frac{\epsilon}{2}\delta_{ij}I  \eea
\bea E_{0}^{ij} = \frac{1}{2}[b_{i}^{+},b_{j}^{+}]_{-\epsilon} = 
\frac{1}{2}(b_{i}^{+}b_{j}^{+}-\epsilon b_{j}^{+}b_{i}^{+}) = 
b_{i}^{+}b_{j}^{+}  \eea
\bea E_{ij}^{0} = \frac{\epsilon}{2}[b_{i},b_{j}]_{-\epsilon} =
\frac{\epsilon}{2}(b_{i}b_{j}-\epsilon b_{j}b_{i}) = 
\epsilon b_{i}b_{j}  \eea
The commutation relations of the
generators (45-47) (pointed out in the Refs.\cite{nish}, \cite{hwa}) are
precisely the commutation relations (20-24) if the following
identifications are made:
\bea A_{ij}=E_{j}^{i},~~  B_{ij}=E_{0}^{ij},~~  C_{ij}=-E_{ij}^{0}  \eea
\vskip 1cm

{\bf The HNN recursive relations}
\vskip .2cm
In the HNN papers series of operators
defined by recursion have been introduced, these series can be
described by the following unique formulas:
\bea ^{0}K^{\alpha}_{\beta}\equiv\delta^{\alpha}_{\beta},~~
^{0}K^{\alpha\beta}\equiv 0,~~^{0}K_{\alpha\beta}\equiv 0 \eea
\bea ^{1}K^{\alpha}_{\beta}\equiv E^{\alpha}_{\beta},~~
^{1}K^{\alpha\beta}\equiv E^{\alpha\beta},~~
^{1}K_{\alpha\beta}\equiv E^{0}_{\alpha\beta}  \eea
\bea ^{m+1}K^{\alpha}_{\beta}\equiv \frac{1}{4}([^{m}K^{\alpha}_{\tau},
E^{\tau}_{\beta}]_{+}+[^{m}K^{\tau}_{\beta},E^{\alpha}_{\tau}]_{+}+
[^{m}K^{\alpha\tau},E^{0}_{\tau\beta}]_{+}+[^{m}K_{\beta\tau},
E_{0}^{\tau\alpha}]_{+})  \eea
\bea ^{m+1}K^{\alpha\beta}\equiv \frac{1}{4}([^{m}K^{\alpha}_{\tau},
E_{0}^{\tau\beta}]_{+}-
\epsilon(-)^{m}[^{m}K^{\beta}_{\tau},E_{0}^{\tau\alpha}]_{+}-
[^{m}K^{\alpha\tau},E^{\beta}_{\tau}]_{+}+[^{m}K^{\tau\beta},E^{\alpha}_{\gamma}
]_{+})  \eea
\bea ^{m+1}K_{\alpha\beta}\equiv \frac{1}{4}([^{m}K^{\tau}_{\alpha},
E^{0}_{\tau\beta}]_{+}-
\epsilon(-)^{m}[^{m}K^{\tau}_{\beta},E^{0}_{\tau\alpha}]_{+}-
[^{m}K_{\alpha\tau},E^{\tau}_{\beta}]_{+}+[^{m}K_{\tau\beta},
E^{\tau}_{\alpha}]_{+}) \eea
with the symmetry properties
\bea ^{m}K^{\alpha\beta}=\epsilon~ ^{m}K^{\beta\alpha},~~
     ^{m}K_{\alpha\beta}=\epsilon~ ^{m}K_{\beta\alpha}~~~
(for~m=even) \eea
\bea ^{m}K^{\alpha\beta}=-\epsilon~ ^{m}K^{\beta\alpha},~~
      ^{m}K_{\alpha\beta}=-\epsilon~ ^{m}K_{\beta\alpha}~~~
(for~m=odd) \eea 

The set of operators $~^{m}K^{\alpha}_{\beta},~^{m}K^{\alpha\beta},~
^{m}K_{\alpha\beta}~$ constructed by recursion in the HNN papers
has tensor properties that depend only on the parity of $m$.
This property, pointed out only by Nishiyama, is however valid
also for the set of operators constructed by Hwa and Nuyts.
\vskip 1cm

{\bf Identification of the kinematical constraints}
\vskip .3cm

 Nishiyama (1976) proved that all kinematical
constraints which intervene in Yamamura's (1974) algebraic
approach to the nuclear collective motion can be obtained from
the following equations of second degree:
\bea ^{2}K^{\alpha}_{\beta}=
\frac{1}{4}(2n-1).\delta^{\alpha}_{\beta},~~~~
^{2}K^{\alpha\beta}= ^{2}K_{\alpha\beta}=0  \eea
Taking into account Eqs.(48) and (50) the last two relations become
\bea ^{2}K^{ij}=2([A_{ik},B_{kj}]_{+}-[B_{ik},A_{jk}]_{+})=
2((AB-BA^{t})+(AB-BA^{t})^{t})_{ij}=0  \eea
\bea ^{2}K_{ij}=2([C_{ik},A_{kj}]_{+}-[A_{ki},C_{kj}]_{+})=
2((CA-A^{t}C)+(CA-A^{t}C)^{t})_{ij}=0  \eea
which are precisely the first two equations (40),(41) written for $\epsilon=+1$
, i.e. for the orthogonal algebra, considered in Nishiyama's paper.

Finally, taking again into account Eqs. (48) and (50), the first
equation becomes
\bea ^{2}K^{i}_{j}=\frac{1}{2}([A_{ik},A_{kj}]_{+}-
[B_{ik},C_{kj}]_{+}) = \frac{1}{4}(2n-1)\delta^{i}_{j}  \eea
whence
\bea A^{2}-BC+((A^{t})^{2}-CB)^{t}=\frac{2n-1}{2}  \eea  
which coincides with Eq. (42).
\vskip 1cm

We shall prove, in the following, that the HNN recursion properties admit a
simple and simultaneous proof. To do that, we note that
the irreducible tensors in $U(L)$ are obtained by symmetrization
from the irreducible tensors in $S(L)$ that transform under the
same irreducible representation; it is therefore sufficient to
prove the recursion properties for the tensors in $S(L)$.
\newpage

\section{Matrix form of the identities}
\vskip .2cm
We start by reminding a procedure imagined by Hannabuss \cite{hann}, and 
extensively applied by Okubo \cite{okubo}, for the derivation of
equations satisfied by the generators of a $given$
irreducible representation (IR) $\lambda$ of a Lie algebra $L$.

Consider a pair $(\lambda,\mu)$ of IRs of $L$ and
the operator
\bea O_{\lambda\mu} = \sum_{i=1}^{dimL}
\lambda(e_{i})\otimes\mu(e^{i}) \eea
where $e_{i}$ are elements of a basis in $L$ and $e^{i}$ are
their dual elements with respect to the Killing-Cartan bilinear form
$(e_{i},e^{j})=\delta_{i}^{j}$.
The operator $O_{\lambda\mu}$ commutes with $\lambda\otimes\mu$
and satisfies \cite{hann} its minimal polynomial:
\bea P(O_{\lambda\mu})=\prod_{\omega\in\Omega(\lambda,\mu)}
[O_{\lambda\mu}-\frac{1}{2}((\omega+2\delta,\omega)-
(\lambda+2\delta,\lambda)-(\mu+2\delta,\mu))I]=0  \eea
where $\Omega(\lambda,\mu)$ is the set of distinct weights in
the Clebsch-Gordan series of the pro-duct $\lambda\otimes\mu$
and $(\lambda+2\delta,\lambda)$ is the eigenvalue of the Casimir
element for the IR $\lambda$.
The degree of the minimal polynomial $P$ is thus equal to the
number of distinct IRs into which $\lambda\otimes\mu$
decomposes. In particular, {\em second-degree} polynomials $P$ 
correspond to Kronecker products $\lambda\otimes\mu$ that
decompose in $two$ irreducible terms.
The matrix elements of the equality $P(O_{\lambda\mu})=0$
with respect to one element of the pair $(\lambda,\mu),~~\mu~$
say, lead to an equation satisfied by the generators of
the IR $\lambda$; the degree of this equation
will be the degree of $P$.

Let us turn back to the equation $~T_{\Lambda}=0~$
associated with the symmetric tensor $~T_{\Lambda}~$ (of highest
weight $\Lambda$) of $U(L)$ derived by projection
(cf. Sec.2) and determine the
IRs $\lambda$ for which $T_{\Lambda}$ vanishes (i.e.
find the modules $\lambda$ annihilated by $T_{\Lambda}$).

We can now look for a convenient "partner" $\mu$ of $\lambda$ such
that the Hannabuss procedure produces an equation 
of second degree  i.e let us find an IR $\mu$ such that 
$\lambda\otimes\mu$ decomposes in two irreducible terms and determine 
this equation. This equation has been proved in \cite{is88-2} to
be precisely $T_{\Lambda}(\lambda)=0$.

 The two ways of finding annihilators (i.e. by projection 
and by the Hannabuss method)
have been proved to be consistent for tensors of second degree
\cite{is88-2}. The combination of both methods
allows to write the set of equations $~T_{\Lambda}=0~$ in a compact and
transparent form, as {\em matrix equations}; these will allow simple
proofs for the recursion relations derived by Hwa and Nuyts \cite{hwa}
and by Nishiyama \cite{nish}. 

As already pointed out at the end of Sec.2, Eqs.(40-42) are
verified for $so(2n)$ (for $sp(2n)$) only if $A,B,C$ are
generators of the IRs $(m\Lambda_{n-1}),(m\Lambda_{n})$
(of the IRs $(m\Lambda_{n})$). It has been proved 
\cite{is88-2} that, for the algebra $so(2n)$, only the Kronecker products
of $(m\Lambda_{n-1})$ and $(m\Lambda_{n})$ with $(\Lambda_{1})$
have Clebsch-Gordan series of length two;
similarly, for $sp(2n)$ only the products
$(m\Lambda_{n})\otimes (\Lambda_{1})$ have two irreducible components:
in both cases there is only one "partner" of the "solutions" of the Eqs.
(40-42): the fundamental representation $(\Lambda_{1})$.

{\bf A classical analogue of the operator}$~O_{\lambda\mu}.$
\vskip .2cm

Let us find now a matrix form for the "classical" relations (35)-(37).

To do that, we replace in Eq.(61) the set of generators of one of the linear
representations of $L$, $\mu(e_{i})$ say, by a set of functions  
that generate a Poisson bracket realisation of $L$ on a
symplectic G-manifold M ($L = Lie(G)$):
\bea e_{i}\in L \rightarrow f_{e_{i}}(p)\in C^{\infty}(M)~~(p\in M) \eea
with the property
\bea f_{Ad(g)e_{i}}(p)=f_{e_{i}}(g^{-1}p)  (g\in G) \eea
Define now the mapping mapping 
\bea  K_{\lambda}: M\rightarrow End V_{\lambda} \eea
(where $~V_{\lambda}~$ is a $\lambda$-module) by
\bea K_{\lambda}:p\in M\longmapsto K_{\lambda}(p)\equiv 
\sum_{i=1}^{dimL} f_{e_{i}}(p)\otimes \lambda(e^{i}) \eea
$K_{\lambda}(p)$ is a matrix whose elements are functions
that generate a Poisson bracket (PB) realisation of $L$.

Denoting by $~g\longmapsto\lambda(g)\in End(V_{\lambda})~$ the
representation of $G$ acting in $V_{\lambda}$, the operator $K_{\lambda}$
is $G$-equivariant; denoting by the same letter $\lambda$ the
representations of $L=Lie(G)$ and of $G$ this means that:
\bea K_{\lambda}(g.p) = \lambda(g^{-1})K_{\lambda}(p)\lambda(g)
(g\in G)  \eea
Any polynomial relation $P(K_{\lambda})=0$ is also $G$-equivariant: it
results that the matrix elements of the polynomial relations
satisfied by $K_{\lambda}$ are polynomial relations satisfied by
the generators of the PB realisation.

As already mentioned, the symmetrisation of these "classical"
polynomial relations gives the "quantum" polynomial relations,
i.e. the relations satisfied by the linear representations. The
elements of both realisations transform as tensors with the same
highest weight and the corresponding extensions of the adjoint
action possesses in both cases the same dimension.
\vskip .5cm
{\bf Identification of the recursive relations}
\vskip .2cm
Let $(\Lambda_{1})$ be the defining representation of
$sp(2n,R)$ or of $o(2n,R)$ and let $~A_{ij},~B_{ij}$ and $C_{ij}$,
$(i,j=1,...,n)~$ be the generators of a PB realisation of these
algebras. The associated $K_{\Lambda_{1}}$-mapping has the expression
$ K_{\Lambda_{1}} = \left( \begin{array}{cc} A_{1} & B_{1} \\
-C_{1} & -A_{1}^{t} \end{array} \right) $
For the algebras $C_{n}$ the matrix $K_{\Lambda_{1}}$ has $N(N+1)/2$
distinct matrix elements; for the algebras $~D_{n},~K_{\Lambda_{1}}~$
has $N(N-1)/2$ distinct matrix elements $~(N=2n)~$. The number
of distinct matrix elements is, for each algebra, equal to the
dimension of the corresponding adjoint representation:
 for $C_{n}$ we have $~dim(ad)=dim(2\Lambda_{1})=\frac{N(N+1)}{2}$
and, for $D_{n}$, $~dim(ad)=dim(\Lambda_{2})=\frac{N(N-1)}{2}$. 

The set of "tensorial identities" (35-38) is equivalent with the
matrix equation

\bea K_{\Lambda_{1}}^{2} = \left( \begin{array}{cc} 
A_{1}^{2}-B_{1}C_{1} & A_{1}B_{1}-B_{1}A_{1}^{t} \\
-C_{1}A_{1}+A_{1}^{t}C_{1} & -C_{1}B_{1}+(A_{1}^{t})^{t} \end{array} \right) 
=\left(\begin{array}{cc}
\frac{I_{2}}{2n} & 0 \\
0 & \frac{I_{2}}{2n} \end{array} \right)\eea

Let us denote    

\bea A_{2}=A_{1}^{2}-B_{1}C_{1},~~B_{2}=A_{1}B_{1}-B_{1}A_{1}^{t},~~
-C_{2}=-C_{1}A_{1}+A_{1}^{t}C_{1} \eea
Reminding Eqs.(7), we get
\bea B_{2}^{t}=\epsilon B_{2},~~C_{2}^{t}=\epsilon C_{2}  \eea
Thus, in contrast to the submatrices $B_{1}$ and $C_{1}$ (cf.(7)),
the submatrices $B_{2}$ and $C_{2}$ are antisymmetric for
the algebras of type $C_{n}$ and symmetric for the algebras of
type $D_{n}$; this phenomenon is general: 
odd powers of $K_{\Lambda_{1}}$
behave like $K_{\Lambda_{1}}$, i.e. their matrix elements are
tensors of type $(2\Lambda_{1})$ for algebras $C_{n}$ and of type
$(\Lambda_{2})$ for algebras $D_{n}$. 

For the $K_{\Lambda_{1}}^{2}$ matrix the submatrices are
antisymmetric for $C_{n}$ algebras and symmetric for $D_{n}$ algebras
(cf. Eqs.(70)). Hence, the number of independent matrix elements
in $K_{\Lambda_{1}}^{2}$ is $N(N-1)/2$ for $C_{n}$ and $N(N+1)/2$
for $D_{n}$. This number of elements of $K_{\Lambda_{1}^{2}}$ is
equal to $dim((0)\oplus(\Lambda_{2})$ for the algebra $C_{n}$
and to $dim((0)\oplus(2\Lambda_{1})$ for the algebra $D_{n}$. Indeed

$dim(0)+dim(\Lambda_{2})=1+\frac{(N+1)(N-2)}{2}=\frac{N(N-1)}{2}
~~~~~   for~~ C_{n}$\\
and 

$dim(0)+dim(2\Lambda_{1})=1+\frac{(N-1)(N+2)}{2}=\frac{N(N+1)}{2}
~~~~~   for~~ D_{n}$\\

For all even powers the matrices behave like $K_{\Lambda_{1}}^{2}$.
 This general statement will be proved now by induction.
\vskip .3cm

{\em Proposition} Let $K_{1}$ be the matrix
\bea K_{1}\equiv\left(\begin{array}{cc} K_{1,A} & K_{1,B} \\
                 K_{1,C} & K_{1,D} \end{array}\right) 
          \equiv\left(\begin{array}{cc} A_{1} & B_{1} \\
                                       -C_{1} & D_{1} \end{array}\right) \eea
whose elements satisfy the relations (7) i.e.
$ D_{1}=-A_{1}^{t},~~~B_{1}=-\epsilon B_{1}^{t},~~~
C_{1}=-\epsilon C_{1}^{t} $

The elements of the matrix 
\bea K_{m}\equiv (K_{1})^{m}\equiv 
\left(\begin{array}{cc}   K_{m,A} & K_{m,B}\\
                         K_{m,C} & K_{m,D} \end{array}\right)\equiv
\left(\begin{array}{cc} A_{m} & B_{m} \\            
                       -C_{m} & -A_{m }^{t} \end{array}\right) \eea   
satisfy the relations
\bea D_{m}=(-1)^{m}A_{m}^{t},~~~ 
     B_{m}=(-1)^{m}\epsilon B_{m}^{t},~~~
     C_{m}=(-1)^{m}\epsilon C_{m}^{t} \eea
\vskip .2cm

{\em Proof.}  Assume that Eqs. (72), (73)  are true. We have to
prove that the elements of the matrix $K_{m+1}=(K_{1})^{m+1}=K_{1}K_{m}=
K_{m}K_{1}$ have the same property.

Because $~A_{1}, B_{1}, C_{1},~ A_{m}, B_{m}, C_{m}~ $ belong to
the symetric algebra, (for each of the algebras under consideration), we have
\bea  K_{m+1}=(K_{1})^{m+1}=K_{1}K_{m}=K_{m}K_{1}  \eea
 
\bea K_{m+1,A}^{t}=(K_{1}K_{m})_{A}^{t}=(K_{m}K_{1})_{A}^{t}=
A_{m}^{t}A_{1}^{t}-C_{m}^{t}B_{1}^{t} \eea 
and 
\bea K_{m+1,D}=(K_{m}K_{1})_{D}=-C_{m}B_{1}+D_{m}D_{1}=
(-1)^{m+1}(A_{m}^{t}A_{1}^{t}-C_{m}^{t}B_{1}^{t} \eea
and the first equality (74) is proved. Similarly
\bea K_{m+1,B}^{t} \equiv (K_{1}K_{m})_{B}^{t}=(A_{1}B_{m}+B_{1}D_{m})^{t}= 
(-1)^{m+1}\epsilon(A_{m}B_{1}-B_{m}A_{1}^{t})     \eea
and
\bea K_{m+1,B}\equiv (K_{m}K_{1})_{B}^{t}=(A_{m}B_{1}+B_{m}D_{1})=
A_{m}B_{1}-B_{m}A_{1}^{t} \eea
this proves the second equality (74). The third equality results in the
same way.


\vskip.3cm
\begin{thebibliography}{99}
\bibitem{yam} M. Yamamura, Prog. Theor. Phys. {\bf 52}, 538 (1974).
\bibitem{nish} S. Nishiyama, Prog. Theor. Phys. {\bf 55}, 1146
(1976).
\bibitem{hwa} R. C. Hwa and J. Nuyts, Phys. Rev. {\bf 145}, 582 (1966).
\bibitem{is85} M. Iosifescu and H. Scutaru, Geometrodynamics Proceedings,
Ed. A. Pr\`astaro, World Scientific (1985), 173.
\bibitem{is88-1} M. Iosifescu and H. Scutaru, Lecture Notes in
Physics, {\bf 313}, Springer (1988), 230.
\bibitem{is88-2} M. Iosifescu and H. Scutaru, J. Math. Phys. {\bf
29}, 742 (1988).
\bibitem{miller} W. Miller, Jr. Symmetry groups and their
applications, Academic Press (1972).
\bibitem{is80} M. Iosifescu and H. Scutaru, J. Math. Phys. {\bf
21}, 2033 (1980).
\bibitem{is86} M. Iosifescu and H. Scutaru, J. Math. Phys.
{\bf 27}, 524 (1986).
\bibitem{s87} H. Scutaru, J. Phys A Math. Gen. {\bf 20}, L1053 (1987)
\bibitem{is90} M. Iosifescu and H. Scutaru, J. Math. Phys. {\bf31},
264 (1990)
\bibitem{hann} K. C. Hannabuss, Characteristic equations for semisimple
Lie groups, Mathematical Institute preprint, Oxford, 1972
\bibitem{okubo} S. Okubo, J. Math. Phys. {\bf 18}, 2382 (1977)

\end{thebibliography}
\end{document}